\documentstyle[manuscript,aps]{revtex}
\begin{document}
\draft
\title{Gas Transport in Porous Media: Simulations and Experiments on
Partially Densified Aerogels}
\author{Anwar Hasmy, Isabelle Beurroies, Daniel Bourret and R\'emi Jullien}
\address{Laboratoire de Science des Mat\'eriaux Vitreux,
Universit\'e Montpellier II, Place Eug\`ene Bataillon,
34095 Montpellier, France}
\date{\today}
\maketitle
\begin{abstract}
The experimental density dependence  of gas (argon and nitrogen) permeability
of partially densified silica	 aerogels in the Knudsen regime
is quantitatively accounted for by a computer model.
The model simulates
both the structure of the sintered material  and the random ballistic motion of
a point
particle inside its voids. The same model is also able to account for the
density dependence
of the specific pore surface as measured
from nitrogen adsorption experiments.
\end{abstract}

\pacs{PACS numbers: 05.60.+w, 66.30.-h, 61.43.-j, 47.55.Mh}

The relation between fluid transport properties in porous materials
and their geometrical characteristics is  the subject of
great experimental and theoretical activities, expecially after the new
insights brought by modern scaling ideas and computer modelling
facilities \cite{1,2}.  In the case of gas transport
in the Knudsen regime,
where the molecular diffusion is
dominated by collisions with the pore walls, it has been demonstrated that
the diffusion properties strongly depend on the geometrical
characteristics of the pore network \cite{3}. Therefore, to properly
account for any experimental results there is a need for a right modelization
of the internal structure of the material on a mesoscopic scale.

In this letter
we  present a comparison between experimental data
and numerical calculations for  the gas permeability
and the specific pore surface of partially densified silica aerogels.
The original (non-densified) aerogel structure is
modelized by the diffusion-limited cluster-cluster aggregation model \cite{4}
which
has been recently shown to successfully explain small angle scattering
experiments on these materials \cite{5}. The densification process is
described by using a numerical
realisation of recent scaling ideas for sintering \cite{6} and the  gas
diffusivity is
calculated  by allowing a point particle to move   within the pores
of the sintered structure along successive random straight lines.

To build the original (non sintered) aerogel structure, we have considered a
three dimensional off-lattice cluster-cluster
model\cite{7,8} which has already been extensively described elsewhere\cite{5}.
A set of $N$ identical spherical particles of diameter $a_0$ are initially
randomly
disposed (without overlap) in a cubic box of edge length $La_0$, occupying a
volume
fraction, or dimension-less concentration of
$c_0 = (\pi / 6) (N / L^3)$.
In the following the particles are assumed to have the density $\rho_S$ (=2.2
g/cm$^3$)
of pure silica so that the actual density is $c_0\rho_S$.
Then, these
particles are allowed to undergo a brownian diffusive motion and they
irreversibly stick when they come at contact. Aggregates of particles
are also able to diffuse together with the individual particles and to
stick to particles or to other aggregates. In this process, the
diffusion constant of the aggregates is considered
to vary as the inverse of their radius of gyration and periodic boundary
conditions are assumed at the box edges. When the concentration is sufficiently
large (larger than a threshold value $c_g$ which tends to zero for infinite
box size\cite{5}), the final structure is
 a set of tangent spheres which can be described
as a loose random packing of connected
fractal aggregates, of fractal dimension $D\simeq 2$,
whose mean size $\xi_0$ decreases as $c_0$ increases.

The sintering process, which consists in a strenghening of the internal
structure accompanied by a gradual elimination of the pores, is here
numerically
simulated by following closely the reasoning of reference\cite{6}. Introducing
a
sintering parameter $s$, which varies from zero (for the non sintered material)
to
a $c_0$-dependent  upper limit (for the compact situation without hole), we
first consider a $s$-
dependent {\it dressed} structure in which each initial sphere of diameter
$a_0$ is replaced by a sphere of same center but of larger diameter
$a_d=a_0(1+s)$.
Then the total volume $V_d(s)$ located inside the overlapping spheres is
numerically
calculated (avoiding
multiple countings of overlaps).

The sintered structure is obtained from the dressed structure after applying a
proper homogeneous volume contraction which insures mass conservation. The
length
reduction factor $\beta(s)$ is given by:

\begin{eqnarray}
\beta(s) = ({6V_d(s)\over N\pi a_0^3})^{1\over 3} =({c(s)\over c_0})^{1\over 3}
\label{E1}
\end{eqnarray}

where

\begin{eqnarray}
c(s) = {V_d(s)\over L^3a_0^3}
\label{E2}
\end{eqnarray}
is the volume fraction of the sintered structure.

In figure 1 we provide    a typical example with $c_0 = 0.1136$. A cut through
the initial
structure is shown as well as at two different stages ($s=1$ and $s=2$) of the
sintering process.
Although in the last stages (when $c(s)$ becomes close to unity)
the shapes of the holes are probably not well reproduced
(they are certainly less smooth than in reality),
this  numerical method gives a quite good variation of the lower
cut-off $a$ as a function of the aerogel density $\rho=c(s)\rho_S$, as it can
be obtained
from low angle neutron scattering experiments on partially densified
silica aerogels. A detailed
presentation of this numerical procedure, comparison with other numerical
methods  and scaling predictions \cite{6} and confrontation with experiments
 will be reported elsewhere\cite{9}.

Here, we would like to focus on another geometrical characteristics which is
measurable
by adsorption experiments, namely the specific pore surface area. Knowing that
the
surface of the dressed structure is simply equal to $2dV_d / da_d =
(2 / a_0) (dV_d(s) / ds)$, the
specific surface $\Sigma(s)$ of the sintered material can be obtained after
correcting by the adequate scaling factor and dividing by the total
mass:

\begin{eqnarray}
\Sigma(s) = {2\over a_0\rho_S}{1\over V_d(s)}{dV_d\over ds}\beta(s) =
{2\over a_0\rho_S}{d\log c(s)\over ds}\beta(s)
\label{E3}
\end{eqnarray}

Starting from a given non-sintered structure, characterized by $L$ and $c_0$,
one can  calculate  $c(s)$ (through $V_d(s)$) as well as its derivative,
and, using formula (3),
one can calculate the specific pore surface area as a function of $s$. Some
results for
the variation of the dimensionless quantity $a_0\rho_S\Sigma(s)$ as a
function of $c(s)$
 are reported
on the log-log plots of figure 2 (dashed curves). One has checked that, for
$s=0$,
one
recovers the exact result
$a_0\rho_S\Sigma(0) = 6$ (horizontal straight line shown in the figure),
which is due to the fact that in that case
the structure is made of identical non overlapping spheres, each
of surface area $\pi a_0^2$ and mass   $\rho_S{\pi\over 6}a_0^3$.
Here the results have been obtained with $L=40.4$ and they result from an
average over 20 realisations of the aerogel structure for each $c_0$ value.
Here and in the following we have checked that the results are almost
independent on $L$. It should be noticed that the
scaling predicted in ref.\cite{6} (a log-log slope of
$D/3(3-D)\simeq -0.7$) is here only observed in a quite small intermediate-$c$
region indicating
that corrections to scaling are important both at low $c$ ($c\sim c_0$) and at
large $c$ ($c\sim 1$).
Obviously $\Sigma$ tends to zero as $c$ tends to one.

To calculate the gas permeability $K(s)$, which is the ratio between a gas flux
through
the sintered material and the applied pressure gradient, we have used the fact
that it is
simply
related to the gas diffusion constant $D_s(s)$ inside the pores by:
\begin{eqnarray}
K(s) = (1-c(s))D_s(s)
\label{E4}
\end{eqnarray}
Introducing the gas diffusion constant inside the pores of the dressed
structure $D_d(s) =
\beta(s)^2D_s(s)$ this formula also writes:
\begin{eqnarray}
K(s) = {1-c(s)\over\beta(s)^2}D_d(s)
\label{E5}
\end{eqnarray}
Since the experimental situation that we would like to describe correspond to
low pressures
and small gas particles (smaller than silica particles), the
diffusion constant $D_d(s)$ is related to the Knudsen motion of a point
particle
inside the pores of the dressed structure. Given the dressed structure,
characterized by
$L$, $c_0$ and $s$, we have
indirectly computed $D_d(s)$ by allowing a hard sphere of diameter $sa_0$
to perform a Knudsen motion inside the voids of the undressed (non sintered)
structure made of spheres of diameter $a_0$. As already noticed \cite{10},
this is strictly equivalent to the original
problem.

Initially a sphere of diameter $sa_0$ is released
at a random point under the condition that it should not overlap any other
sphere (when
$s$ and/or $c_0$ are too large, many trials are necessary)
and a random direction (uniformly distributed in space)
is selected. Then the
moving sphere is allowed to follow a random straight line motion in this
direction until it collides
a sphere of diameter $a_0$. Immediately after collision a new random direction
is selected
(in a half space) according to the so-called Knudsen cosine law \cite{3,11}, i.
e.  such that its
probability distribution is proportional to $\cos\theta d^2\Omega$ where
$\theta$
is the angle
between this direction and the normal (center-to-center direction) and where
$d^2\Omega = \sin\theta d\theta d\phi$ is the elementary solid angle. In
practice $\theta$ is chosen such that
$\cos^2\theta$ is uniformly distributed between zero and one  while the
azimutal angle
is chosen  uniformly between 0 and $2\pi$.  It has been shown that
this is essential to recover the equilibrium Boltzmann statistics and the
right transport coefficients \cite{3,11}. Then, after a large
number $N_c$ of collisions and using the periodic boundary conditions,
one calculates the end-to-end square
displacement $\ell^2$ of the diffusing sphere center as well as the total
legth $\Lambda$ of its trajectory, which is the sum of the lengths of the
successive
segments. This allows us to calculate $D_d(s)$ by the well-known formula:
\begin{eqnarray}
D_d(s) = {1\over 6} v {\ell^2\over\Lambda}
\label{E6}
\end{eqnarray}
where $v=\sqrt{8kT / \pi m}$ is the mean molecular velocity of the gas particle
of
mass $m$ at temperature $T$. Reporting in formula (5), one finally gets:
\begin{eqnarray}
K(s) = {1\over 6} v{1-c(s)\over\beta(s)^2}{\ell^2\over\Lambda}
\label{E7}
\end{eqnarray}
Some results for the variation with $c(s)$ of the dimensionless quantity
 $K(s) / a_0v$ are reported in the log-log plot of figure 3. In such
calculation
we have taken $\Lambda\simeq 10L$ and we have averaged over 10 realizations of
the
original structure as well as 60 trials for the initial position,
for each value of $c_0$. As expected
$K$ tends to zero as $c$ tends to one.
The curves obtained for different $c_0$ are quite close to each other, showing
that
the density dependence of the permeability is not strongly sensitive to the
details of
the structure (tangent or overlaping spheres). However we have shown above
(see figure 2) that considering
overlaping spheres to simulate the sintering is essential to account for a
decrease
of the specific pore surface with increasing densities.

Our simulation also allows us to investigate the statistics of the chords,
which are the segments between successive collisions. It is known that
the chord length distribution enters
many properties of a porous material \cite{3,12}. In particular
the mean chord length $<\ell>$ is directly related to the specific pore surface
via
a well known mathematical theorem \cite{13}.  In our simulation the mean chord
length
of the dressed structure
can be simply obtained by
$<\ell> = {\Lambda\over N_c}$.
Then going back to the sintered structure and using the theorem,
we can
derive the following alternative expression for $\Sigma(s)$:
\begin{eqnarray}
\Sigma(s) = 4{1-c(s)\over c(s)}{\beta(s)\over\rho_S<\ell>}
\label{E8}
\end{eqnarray}
We have reported $a_0\rho_S\Sigma$ as a function of $c(s)$, calculated from
$<\ell>$ with the
same $c_0$ values as before on the same plot of figure 2. While the overall
$c$-dependence
is qualitatively the same, we observe some discrepancies compared to the
preceeding calculation
which increase for larger $c_0$. Even if we have checked that the discrepancies
do not result
from introducing an upper cut off to the chord length at $\Lambda$,
the numerical uncertainties are certainly
larger in the second calculation. It may also happen that, for larger
concentrations, the presence
of closed pores may affect differently the two calculations (closed pore
surface is fully
considered in the first calculation). Moreover, the theorem holds for a
globally isotropic
pore network and, even if all the cluster-cluster growth conditions are
isotropic,
the resulting gel may be not since it is known that the cluster-cluster
aggregates are intrinsically anisotropic \cite{14}.

The gas permeability experiments on partially densified aerogels have been
performed
using a newly designed apparatus which allows to measures simultaneously the
gas flux through
a cylindrical sample and the gas pressure
difference between the two edges. Technical details will be given elsewhere
\cite{15}.
Here we report on experiments done with a series of aerogels which have been
prepared
under neutral conditions and for which pore surface area data have been
previously
measured by nitrogen adsorption experiments \cite{6}. Before sintering, their
density
is known to be equal to $\rho_0 = 0.25$ g/cm$^3$. Two series of experiments
have been done
with different gas, nitrogen and argon. The experimental results for both the
ratio $K / v$
and the pore surface area $\Sigma$ are reported in figure 4. As expected,
one observes that, after
dividing by the gas velocity, the data are superimposed. In
both cases the symbols correspond to the experimental data and the full lines
correspond
to a fit with our computer model using $c_0 = \rho_0 / \rho_S = 0.1136$,
$L=40.4$, and
$a_0= 70\AA\ $. It is remarkable that, with only one adjustable parameter
($a_0$), we
are able to fit quite well the two quantities simultaneously (formula (3) has
been used for $\Sigma$).
This gives a good confidence in the
pertinence of our model. Although of the right order of magnitude,
the value chosen for $a_0$ is larger than
other previous estimates for the mean particle diameter of neutral
aerogels\cite{16}. Here,
we have considered a set of monodisperse spheres while
it is well known that the size polydispersity of individual particles is very
large
in neutral aerogels\cite{16}. As already noticed for small angle scattering
experiments\cite{17}, the large
balls dominate any scattering process and they also have a greater contribution
to
the internal pore surface area.  This could explain why we are
obliged to consider here an effective particle diameter larger than the simple
mean. Also, counting the
closed pores (which are consisered in the fit) could also increase the
effective $a_0$.

In conclusion, using a computer model which simulates both the  formation and
densification process of aerogels we have been able to quantitatively account
for the density dependence of the gas permeability and pore surface area. In
the future
we will pursue the numerical simulations of aerogels sintering, and we will
extend the present model, introducing polydispersity and analysing
more extensively the chord length distribution. We have also in project to
simulate the brownian diffusion regime in connection with the transport
properties
of fluids inside aerogels.

We would like to acknowledge discussions with R. Sempere, T. Woignier, J.
Phalippou,
M. Foret and J. Pelous. One of us (A. H.) would like to acknowledge support
from
CONICIT(Venezuela).
Laboratoire de Sciences des Mat\'eriaux Vitreux is Unit\'e Associ\'ee au
CNRS No. 1119.

\begin{figure}
\caption{Cut of the simulated aerogel structure at three stages of sintering,
$s=0, 1$ and 2, where $c(s)= 0.1136, 0.54, 0.88$, repectively. Here and
in the other figures the box size is $L=40.4$.}
\end{figure}

\begin{figure}
\caption{Log-log plot of $a_0\rho_S\Sigma$ versus $c$. Dotted, dashed and
dot-dashed curves
correspond to formula (3) with $c_0= 0.05, 0.1136$ and 0.2, respectively. The
horizontal line
correspond to the exact result $a_0\rho_S\Sigma = 6$ for $c=c_0$. Triangles,
diamonds and
squares correspond to formula (8) with $c_0= 0.05, 0.1136$ and 0.2,
respectively.}
\end{figure}

\begin{figure}
\caption{Log-log plot of $K / a_0v$ versus $c$. Triangles, diamonds and squares
correspond to $c_0 = 0.05, 0.1136$ and 0.2, respectively.}
\end{figure}

\begin{figure}
\caption{Fits of both $K / v$ and $\Sigma$ for neutral aerogels.
In the case of $K / v$ open squares and diamonds correspond to argon and
nitrogen, respectively. The lines are the fits
with $L=40.4$, $c_0=0.1136$ and $a_0= 70\AA\ $.}
\end{figure}
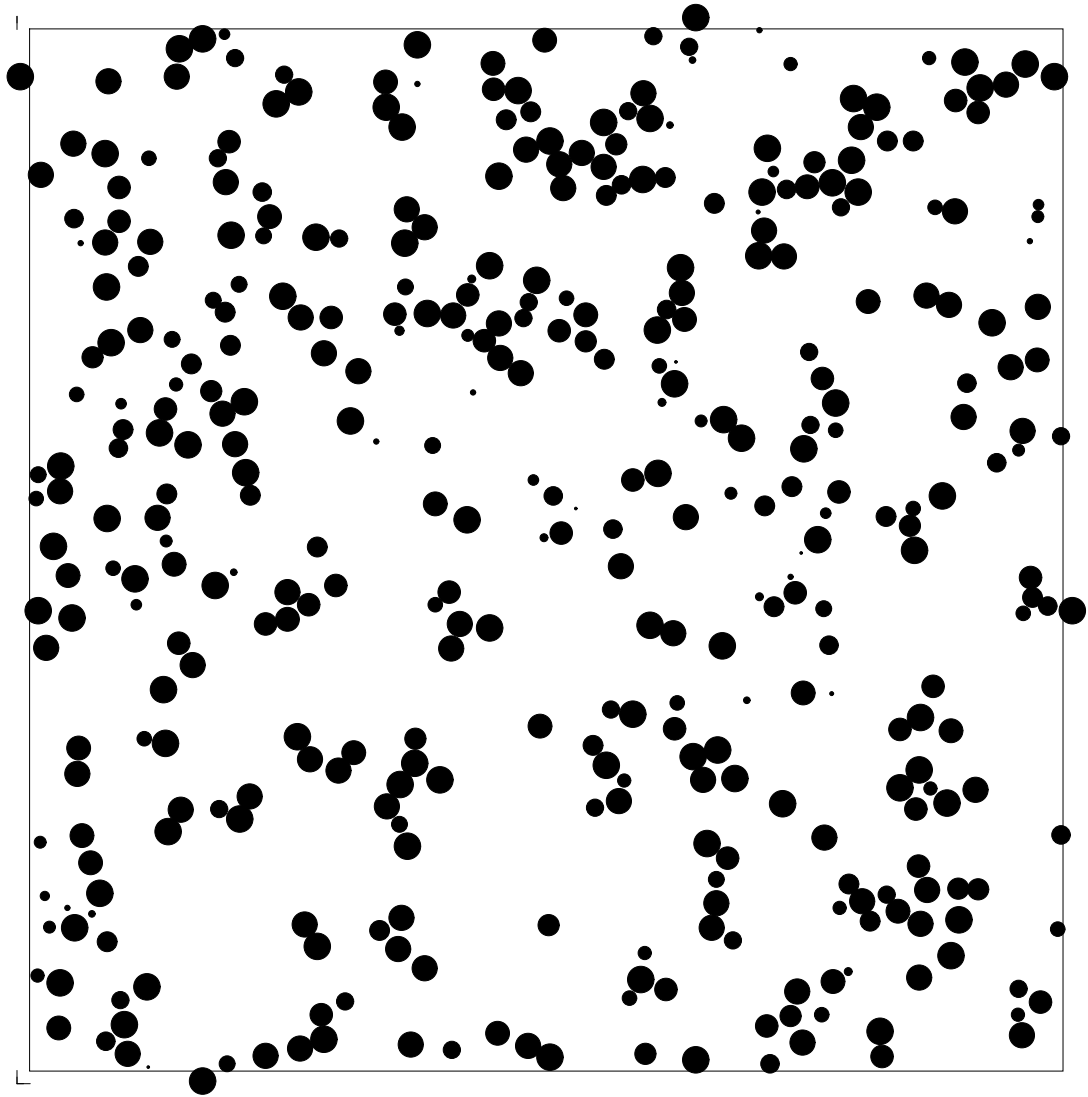
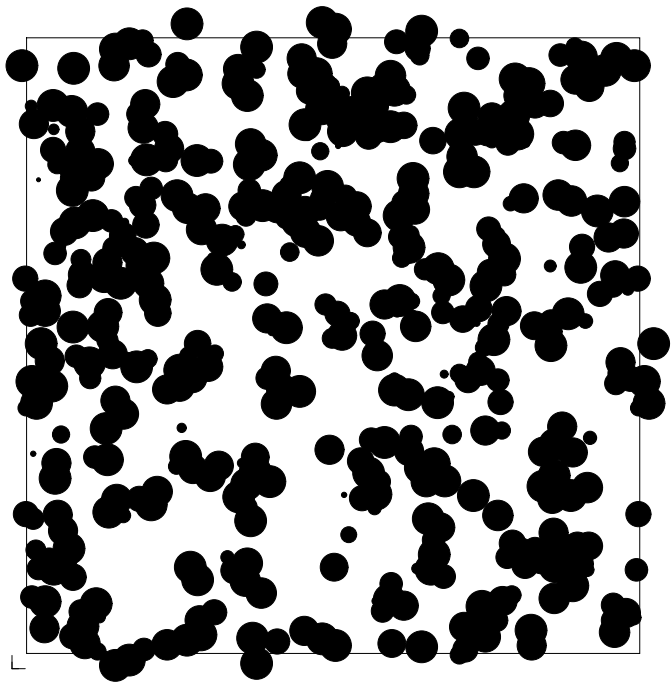
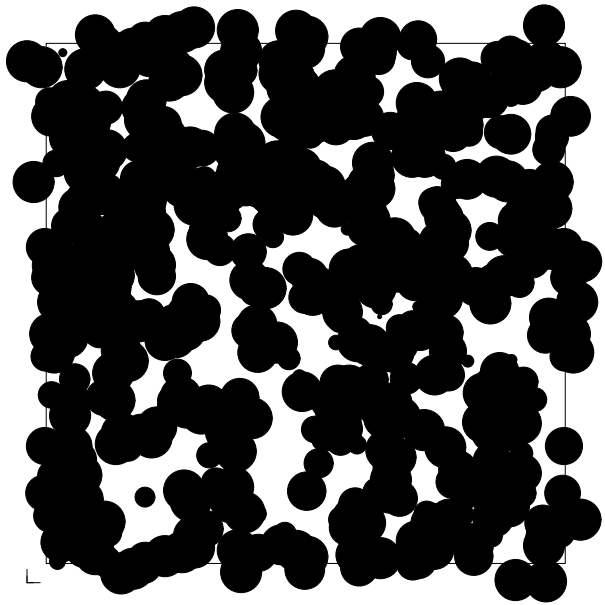
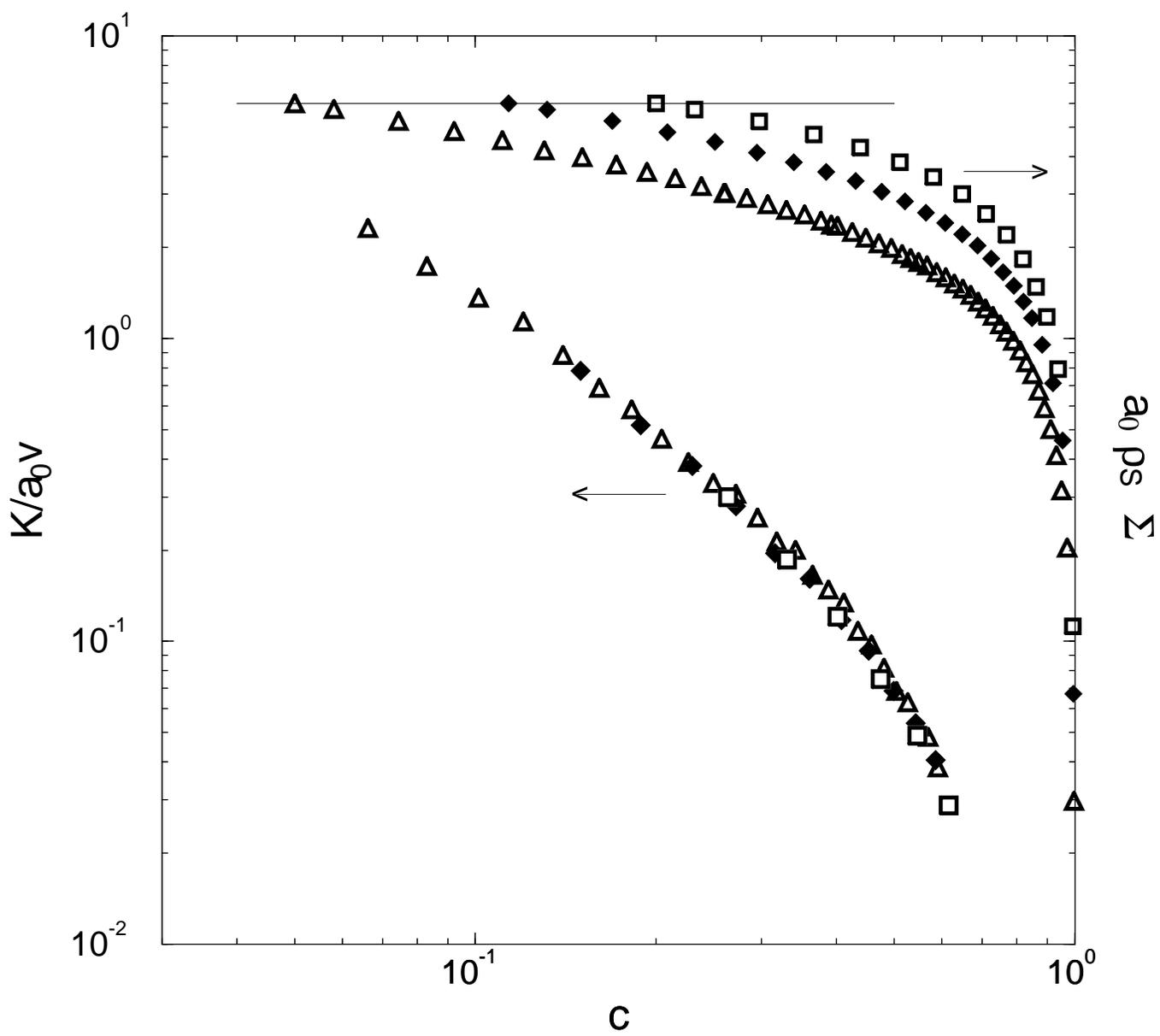
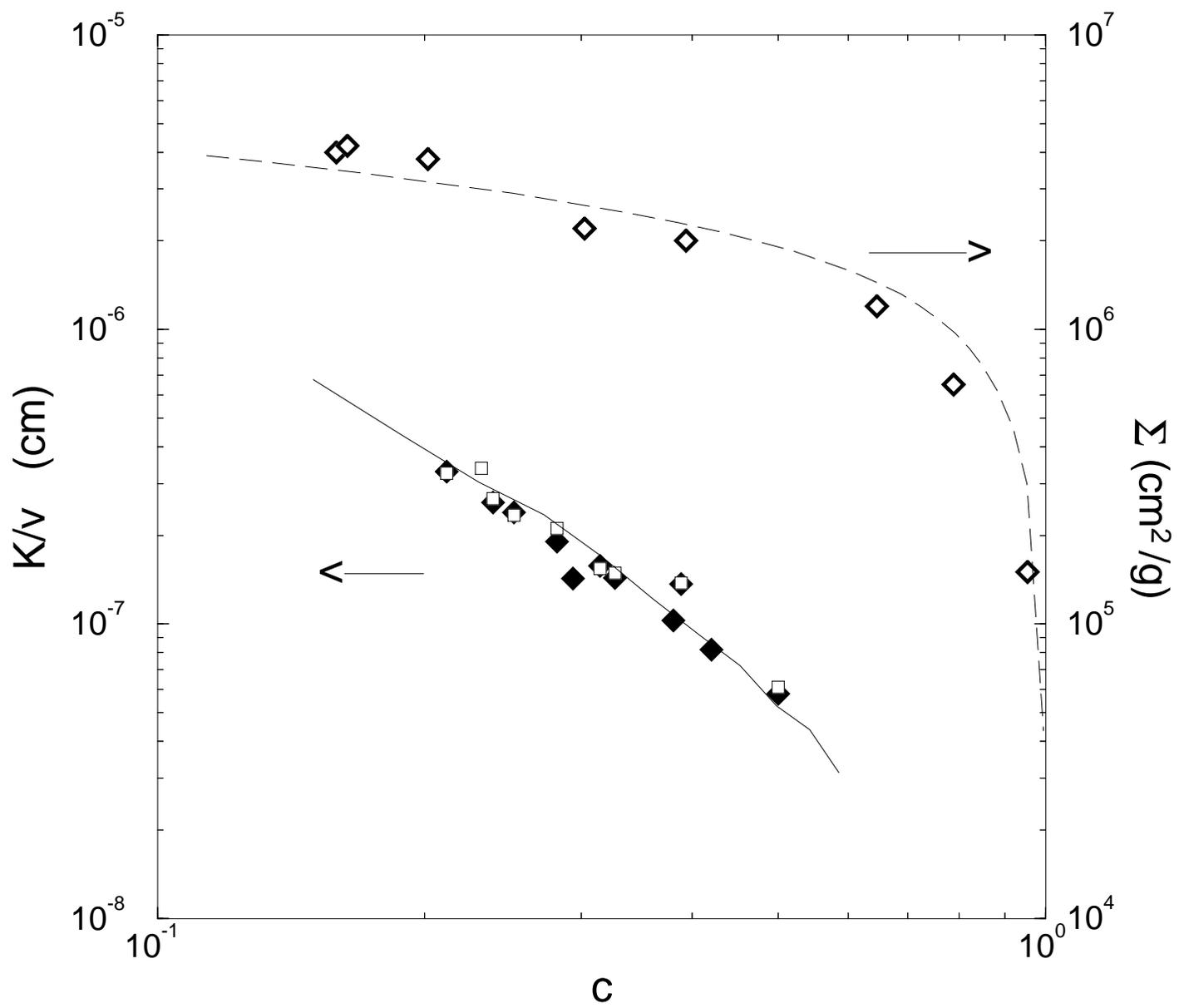
\end{document}